# Reversibility of dynamics and multiple-quantum coherences

A. K. Khitrin

*Department of Chemistry, Kent State University, Kent, OH 44242, USA*

In spin systems, the decay of the Loschmidt echo in the time-reversal experiment (evolution – perturbation – time-reversed evolution) is linked to the generation of multiple-quantum coherences. The approach is extended to other systems, and the general problem of reversibility of quantum dynamics is analyzed.

## I. INTRODUCTION

Time-reversed evolution can be achieved by changing a sign of the Hamiltonian. The echo, resulting from the forward – backward evolution of an isolated system, is often called the Loschmidt echo, in relation to the Loschmidt paradox. The first experimental implementation of such two-way evolution for a many-body system ("magic echo") was done by Waugh and colleagues [1] for a system of dipolar-coupled nuclear spins of a solid. The scaled Hamiltonian of dipolar interactions with reversed sign has been created approximately, as the average Hamiltonian [2], generated by a sequence of radio-frequency pulses in an nuclear magnetic resonance (NMR) experiment. The echo has been observed at times about ten times longer than the characteristic time of spin dynamics $T_2$. It is possible to extend the echo decay time by about one order of magnitude by using the average Hamiltonians for both the forward and backward evolutions [3].

The decay of the Loschmidt echo results from a non-perfect reversal of the Hamiltonian $H$ or from the coupling of the system of interest to the rest of the universe. As an example, one may assume that the Hamiltonian during the time-reversed evolution is not $H_{rev} = -H$ but $H_{rev} = -UHU^{-1}$, where $U$ is some unitary transformation close to identity. Equivalently, one may apply the same perturbation $U$ to the state of the system after the forward evolution. Therefore, the general scheme of the time-reversal experiment can be viewed as evolution – perturbation – exact time-reversed evolution.

In classical mechanics, two initially close points in a phase space, representing the states of a system, can exponentially diverge at a later moment of time $t$, so that the distance between the points increases as $exp(\lambda t)$, where $\lambda$ is the largest Lyapunov exponent. Such dynamics is called mixing. Mixing and ergodicity are two important properties of dynamics which justify application of statistical methods and thermodynamics. Large change of a state caused by a small perturbation in the past is often called the "butterfly effect" [4]. Dynamics of a classical system with only few degrees of freedom can be mixing. The Sinai billiard is an example with two degrees of freedom. For the Loschmidt echo, with the evolution time $\tau$ in one direction, one would expect that its amplitude $M(2\tau)$ exponentially deviates from the ideal echo: $\delta M = M(0) - M(2\tau) \propto exp(\lambda\tau)$, where $\lambda$ is on the order of the largest Lyapunov exponent.

From the correspondence principle, one may expect a similar behavior of large quantum systems. Quantum evolution is unitary and preserves the distances between the states. Therefore, for the comparison, the values of observables rather than the distances between the states should be used. As one example of expected similarity between the quantum and classical dynamics, it has been shown that the dynamics of a lattice of "classical spins" (precessing magnetic moments) represents very closely the dynamics of interacting spins ½ in a process of spin diffusion [5]. Similar to classical, in quantum systems with time-independent Hamiltonians, small initial perturbations can also be amplified by a subsequent dynamic evolution. The example is the exactly solvable "quantum domino" dynamics in a spin chain [6], where one initially flipped spin causes a reversal of magnetization of the entire cluster. The magnetization change is linear in time for this model. The schemes of amplified quantum measurement [7] and "quantum butterfly effect" [8] have been demonstrated experimentally for spin clusters. However, as we will see below, the behavior of quantum systems in the Loschmidt echo experiment and the whole concept of reversibility are very different from those in classical systems.

In a recent simulation [9] a comparison has been made between the systems of "classical spins" and spins ½ in the Loschmidt echo experiment. While the classical system demonstrated an exponential growth of δ$M$, consistent with the estimated value of the Lyapunov exponent, the quantum system (5 x 5 lattice of spins ½ with periodic boundary conditions) showed strongly non-exponential behavior. As we will see below, the effect of small perturbations on quantum systems in a time-reversal experiment can be analyzed quantitatively. We will also see that the reversibility of quantum dynamics depends not only on the main Hamiltonian, but also on the observable of interest and the perturbation Hamiltonian. We will start with the spin systems and then generalize the results when possible.

## II. HAMILTONIAN

Coupled nuclear spins ½ in solids are still the most suitable systems for the experimental exploration of the time-reversed dynamics. In solids, spin degrees of freedom can be extremely well isolated from other degrees of freedom (lattice) and spin Hamiltonians can be modified by applying sequences of radio-frequency pulses to create the desired average Hamiltonians. We will consider spin Hamiltonians of the form:

$$H = \sum_{i>j} b_{ij} \left( a S_{xi} S_{xj} + b S_{yi} S_{yj} + c S_{zi} S_{zj} \right). \qquad (1)$$

As an example, for the secular part of the dipole-dipole interaction in strong z-field $H_{dz}$, one has $a = b = 1$, $c = -2$, and the coupling constants $b_{ij} \propto (3\cos^2\theta-1) r_{ij}^{-3}$, where $r_{ij}$ is the distance between spins $i$ and $j$, and θ is the angle between $r_{ij}$ and z-axis. Pure double-quantum average Hamiltonian (with respect to x-axis), created by the pulse sequence in ref. [3], is described by a = 0, b = 1, and c = -1 (here and below $x$ is chosen to be the quantization axis). The general Hamiltonian (1) can be decomposed into five terms of different symmetry with respect to the rotation about x-axis:

$$H = \sum_{n=-2}^{n=2} H_n, \quad [S_x, H_n] = n H_n, \quad \text{or} \quad e^{i\varphi S_x} H_n e^{-i\varphi S_x} = e^{in\varphi} H_n. \qquad (2)$$

$H_n + H_{-n}$ can be called the *n*-quantum Hamiltonian.

## III. MULTIPLE-QUANTUM COHERENCES

The equation of motion for the density matrix ρ($t$) is

$$\frac{d}{dt}\rho(t) = -i[H, \rho(t)], \qquad (3)$$

and the time evolution is

$$\rho(t) = e^{-iHt}\rho(0)e^{iHt} = e^{Lt}\rho(0) = \sum_{n=0}^{\infty}\frac{t^n}{n!} L^n \rho(0), \qquad (4)$$

where ρ(0) is the initial density matrix and $L = -i[H, ...]$ is the Liouvillian. If one starts with the density matrix ρ(0), which is invariant under x-rotations: $[S_x, \rho(0)] = 0$, then ρ(t) at any given moment t can be decomposed into terms of different symmetry with respect to x-rotations:

$$\rho(t) = \sum_n \rho_n(t), \tag{5}$$

$$[S_x, \rho_n(t)] = n\rho_n(t), \quad \text{or} \quad e^{i\varphi S_x}\rho_n(t)e^{-i\varphi S_x} = e^{in\varphi}\rho_n(t). \tag{6}$$

The term $\rho_n(t)$ is called the *n*-quantum coherence [10]. Multiple-quantum (MQ) coherences $\rho_n(t)$ can be viewed as the Fourier components of the density matrix ρ(t) transformed by the x-rotation. The normalized intensities of the *n*-quantum coherences are defined as

$$I_n(t) = Tr\{\rho_n(t)\rho_{-n}(t)\}/Tr\{\rho(0)^2\}. \tag{7}$$

$Tr\{\rho(t)^2\}$ does not depend on time:

$$\frac{d}{dt}Tr\{\rho(t)^2\} = 2Tr\{-i[H,\rho(t)]\rho(t)\} = -2iTr\{H[\rho(t),\rho(t)]\} = 0. \tag{8}$$

Since $Tr\{\rho(t)^2\} = Tr\{\rho(0)^2\}$ and $Tr\{\rho_n\rho_m\} = 0$ when $m \neq -n$ (traces are invariant under rotations), $Tr\{\rho(t)^2\} = Tr\{\sum_n \rho_n(t)\rho_{-n}(t)\} = Tr\{\rho(0)^2\}$, and the sum of intensities of the multiple-quantum (MQ) coherences is conserved: $\sum_n I_n(t) = 1$. It is also obvious that $I_n = I_{-n}$.

Experimentally, the intensities of the MQ coherences can be measured by either converting the coherences back to magnetization using a time-reversed evolution [11-13,3] or directly, by performing a projective quantum measurement [14]. It is also possible to selectively excite the MQ coherences of desired orders [15].

MQ coherences give a convenient, but incomplete, description of spin correlations. *N*-spin correlation is represented by a term in the density matrix which is a product of *N* single-spin operators. For $n > 0$, *n*Q coherence which contains the smallest number of correlated spins has the form $S^+_1 S^+_2 S^+_3 \ldots S^+_n$ where

$$S_i^\pm = S_{yi} \pm iS_{zi}. \tag{9}$$

Therefore, *n*Q coherence can appear only when at least $|n|$ spins are correlated. A uniform x-rotation of all spins by the angle φ results in the added phase *n*φ for the *n*Q coherence as in eq.(6). It has been proposed to use this high sensitivity of the MQ coherences to rotations in spectroscopy and high-precision frequency measurements [16]. Filtering of the MQ coherences has been used to create pseudo-pure states in clusters of up to twelve nuclear spins [17]. Measurement of the MQ intensities gives an experimental method of studying multi-spin correlations of very high orders [18]. nQ coherences with $n \approx 100$ has been detected [19]. Within the statistical approach [3], it requires a correlation of about $n^2 \approx 10^4$ spins.

## IV. LOSCHMIDT ECHO

For now, to be specific, we assume that the initial high-temperature state is described by the density matrix $\rho(0) = \rho_0(0) = S_x$, the forward and backward evolutions, with durations τ each, are governed by the Hamiltonians *H* and *-H*, respectively, and that the perturbation of the state after the forward evolution is a uniform rotation of all spins by a small angle δ around x-axis. The measurable quantity is the x-component of the total magnetization $M_x(t) = Tr\{S_x \rho(t)\}$. Then, the normalized amplitude of the Loschmidt echo at the moment 2τ can be written as

$$M_x(2\tau) = Tr\{S_x e^{iH\tau} e^{i\delta S_x} e^{-iH\tau} S_x e^{iH\tau} e^{-i\delta S_x} e^{-iH\tau}\}/Tr\{S_x^2\} \tag{10a}$$

$$= Tr\{e^{-iH\tau} S_x e^{iH\tau} e^{i\delta S_x} e^{-iH\tau} S_x e^{iH\tau} e^{-i\delta S_x}\}/Tr\{S_x^2\} \tag{10b}$$

$$= Tr\{\rho(\tau) e^{i\delta S_x} \rho(\tau) e^{-i\delta S_x}\}/Tr\{S_x^2\} = Tr\{\sum_n \rho_n(\tau) \sum_m \rho_m(\tau) e^{im\delta}\}/Tr\{S_x^2\} \tag{10c}$$

$$= Tr\{\sum_n \rho_{-n}(\tau) \rho_n(\tau) e^{in\delta}\}/Tr\{S_x^2\} = \sum_n I_n(\tau) e^{in\delta}. \tag{10d}$$

We are interested in the behavior of the Loschmidt echo $M_x(2\tau)$ at finite values of $\tau$ and in the limit of small perturbation $\delta \to 0$. Since $I_n = I_{-n}$, we find from eq.(10d) that $dM_x(2\tau)/d\delta = 0$. The second derivative is

$$\frac{d^2 M_x(2\tau)}{d\delta^2} = -\sum_n n^2 I_n(\tau) = -m_2(\tau), \tag{11}$$

where $m_2(\tau)$ is the second moment of the distribution of normalized MQ intensities. Therefore, the echo amplitude is

$$M_x(2\tau) = 1 - \frac{1}{2}\delta^2 m_2(\tau), \tag{12}$$

and the decay $\delta M$ of the Loschmidt echo is

$$\delta M_x(\tau) = M_x(0) - M_x(2\tau) = \frac{1}{2}\delta^2 m_2(\tau). \tag{13}$$

One can see that irreversibility of the dynamics requires an unlimited growth of the width of the MQ intensities distribution: $m_2(\tau) \to \infty$ at $\tau \to \infty$. Such behavior can be viewed as the quantum analog of the mixing dynamics. The unlimited growth of $m_2(\tau)$ also means an unlimited growth of the spin correlation order (the number of correlated spins). However, as we will see below, the reversed statement is not true. Unrestricted growth of spin correlations does not guarantee an unlimited growth of $m_2(\tau)$ and, therefore, does not necessary lead to the irreversible dynamics.

## V. GENERALIZATION

In eq.(10a) we will replace the special initial condition $\rho(0) = S_x$ by a general initial condition $\rho(0)$, and the generator of rotations $S_x$ by an arbitrary operator $V$:

$$M(2\tau) = Tr\{\rho(0) e^{iH\tau} e^{iV\delta} e^{-iH\tau} \rho(0) e^{iH\tau} e^{-iV\delta} e^{-iH\tau}\}/Tr\{\rho(0)^2\} \tag{14a}$$

$$= Tr\{\rho(\tau) e^{iV\delta} \rho(\tau) e^{-iV\delta}\}/Tr\{\rho(0)^2\}. \tag{14b}$$

The perturbation of the density matrix at the moment $\tau$ can now be viewed as caused by the Hamiltonian $-V$, which acts during the time interval $\delta$. $M(2\tau)$ in eq.(14a) now has a meaning of an overlap between the initial density matrix $\rho(0)$ and the density matrix at the end of the time-reversed evolution $\rho(2\tau)$. Similar to eqs.(5,6), we can introduce the Fourier components $\rho_\omega(t)$ of the density matrix $\rho(t)$, with respect to $\delta$, after the transformation $e^{iV\delta} \rho(t) e^{-iV\delta}$:

$$\rho(t) = \int d\omega\, \rho_\omega(t), \tag{15}$$

$$[V, \rho_\omega(t)] = \omega \rho_\omega(t), \quad \text{or} \quad e^{iV\delta} \rho_\omega(t) e^{-iV\delta} = e^{i\omega\delta} \rho_\omega(t). \tag{16}$$

The term "MQ coherence" is not meaningful for the component $\rho_\omega(t)$, so we can call it the "V-coherence" to emphasize that the transformation properties are defined with respect to the perturbation *V*. In exactly the same way as it has been done for the discrete case in Section IV, by introducing

$$I_\omega(t) = Tr\{\rho_\omega(t)\rho_{-\omega}(t)\}/Tr\{\rho(0)^2\} \tag{17}$$

and

$$m_2(t) = \int d\omega \; \omega^2 \, I_\omega(t), \tag{18}$$

one obtains the same eq.(13) for the decay of the Loschmidt echo.

## VI. WEAK IRREVERSIBILITY

An alternative way of calculating the echo decay is to apply the perturbation to the Hamiltonian, rather than to the density matrix $\rho(\tau)$. An equivalent form of eq.(10b) is

$$M_x(2\tau) = Tr\{e^{-iH\tau}S_x e^{iH\tau} e^{-i(H+H')\tau} S_x e^{i(H+H')\tau}\}/Tr\{S_x^2\} \tag{19}$$

where

$$H' = i\delta \, [S_x, H] = i\delta \sum_{n=-2}^{n=2} n H_n \,. \tag{20}$$

In the interaction frame (which eliminates the main Hamiltonian *H*), eq.(19) reduces to

$$M_x(2\tau) = Tr\{S_x \, \tilde{\rho}(\tau)\}/Tr\{S_x^2\}, \tag{21}$$

where

$$\frac{d}{dt}\tilde{\rho}(t) = -i[\tilde{H}'(t), \tilde{\rho}(t)], \qquad \tilde{H}'(t) = e^{iHt} H' e^{-iHt} \,. \tag{22}$$

The solution to $\tilde{\rho}(t)$ can be obtained by iterations as $\tilde{\rho}(t) = S_x + \tilde{\rho}^{(1)} + \tilde{\rho}^{(2)} + \cdots$, where $\tilde{\rho}^{(1)}$ does not contribute to eq.(21) and

$$\tilde{\rho}^{(2)}(t) = -\int_0^t dt' \int_0^{t'} dt'' \, [\tilde{H}'(t'), [\tilde{H}'(t''), S_x]]. \tag{23}$$

Therefore,

$$M_x(2\tau) = 1 - \int_0^\tau dt' \int_0^{t'} dt'' \, \frac{Tr\{S_x[\tilde{H}'(t'), [\tilde{H}'(t''), S_x]]\}}{Tr\{S_x^2\}}$$

$$= 1 + \int_0^\tau dt' \int_0^{t'} dt'' \, \frac{Tr\{[S_x, \tilde{H}'(t')][S_x, \tilde{H}'(t'')]\}}{Tr\{S_x^2\}}$$

$$= 1 - \sum_{n=-2}^{n=2} \delta^2 n^2 \int_0^\tau dt' \int_0^{t'} dt'' \, \frac{Tr\{\tilde{H}_n(t')\tilde{H}_{-n}(t'')\}}{Tr\{S_x^2\}}, \tag{24}$$

where we used eq.(20) for the perturbation Hamiltonian. By introducing the correlation functions

$$g_n(t) = \frac{Tr\{\tilde{H}_n(t)\tilde{H}_{-n}(0)\}}{Tr\{S_x^2\}} \tag{25}$$

and the correlation times

$$\tau_n = \frac{1}{2}\int_{-\infty}^{\infty} dt\, g_n(t)/g_n(0), \tag{26}$$

one can obtain from eq.(24) at $\tau \gg \tau_n$

$$M_x(2\tau) = 1 - \tau\delta^2 \sum_{n=-2}^{n=2} n^2\, \tau_n\, Tr\{H_n H_{-n}\}/Tr\{S_x^2\}. \tag{27}$$

The deviation of the Loschmidt echo grows very slowly, as a linear function of time. We will call such behavior with $\delta M(\tau) \propto \tau$ the weak irreversibility. The necessary condition for the weak irreversibility is the existence of the correlation times (26), i. e. the correlation functions (25) should decay faster than $t^{-1}$. A comparison between eqs.(27) and (12) gives

$$m_2(\tau) = \tau \sum_{n=-2}^{n=2} 2n^2\, \tau_n\, Tr\{H_n H_{-n}\}/Tr\{S_x^2\}. \tag{28}$$

We see that the weak irreversibility also means that the second moment of the MQ intensities distribution grows linearly with time at $t \gg \tau_n$. Such linear growth at long times has been already observed in the early experiments [3, Fig.7] (MQ intensities were fitted by a Gaussian, and the growth of its variance has been reported). For the Hamiltonian $H_{dz}$ ($a = 1$, $b = 1$, $c = -2$) and the yy-zz Hamiltonian ($a = 0$, $b = 1$, $c = -1$), used in most of the experiments, there are only terms with $|n| = 2$ in eqs. (27) and (28). It is interesting that shorter correlation times $\tau_n$ of the "transverse" correlation functions, which would normally be viewed as shorter dynamic memory, cause better reversibility of $S_x$.

For an estimate, we will introduce the strength of the local fields $\omega_{loc}^2 = Tr\{H^2\}/Tr\{S_x^2\}$ and replace $\tau_n$ by a single correlation time $\tau_c \approx \omega_{loc}^{-1}$. The approximate asymptotic expressions for $\delta M(\tau)$ and $m_2(\tau)$ are

$$\delta M(\tau) \approx \tau\, \delta^2\, \omega_{loc} \quad \text{and} \quad m_2(\tau) \approx \tau\, \omega_{loc}. \tag{29}$$

The linear growth of $m_2(\tau)$ is consistent with a diffusion or "random walk" [3,20] in a space where the coordinate is the MQ coherence order. It should be noted, however, that MQ dynamics is fully reversible, and cannot be adequately described by a random process.

For an arbitrary perturbation $V$, the main Hamiltonian $H$ can be decomposed into the harmonics:

$$H = \int d\omega\, H_\omega, \qquad [V, H_\omega] = \omega\, H_\omega. \tag{30}$$

The continuous version of eq.(27) will be

$$\delta M_x(2\tau) = \tau\delta^2 \int d\omega\, \omega^2\, \tau_\omega\, Tr\{H_\omega H_{-\omega}\}/Tr\{S_x^2\}. \tag{31}$$

Therefore, the necessary condition of weak irreversibility is the convergence of the integral in eq.(31)

$$\int d\omega\, \omega^2\, \tau_\omega\, Tr\{H_\omega H_{-\omega}\}/Tr\{S_x^2\} < \infty. \tag{32}$$

We can note that the convergence depends indirectly on the spectrum of the perturbation. As an example, for the uniform rotation, the spectrum of $S_x$ at $N \to \infty$ is unlimited, but the selection rules leave only few harmonics $H_n$, and the integral in eq.(32) is finite when the correlation times $\tau_n$ exist. A different situation is expected when the perturbation is an interaction with the lattice (the coupling constants $b_{ij}$ in the Hamiltonian (1) should be viewed as the operators in this case). The frequency spectrum of nuclear motions is virtually unlimited compared to the frequencies of nuclear spin motion, the condition (32) is violated, and we do not expect weak irreversibility.

## VII. EXACTLY SOLVABLE MODELS

There are two known spin models where the evolution of the MQ coherences can be calculated exactly.

*a) zz model.*

For the Hamiltonian (1) with $a = b = 0$, and $c = 1$ spin dynamics simplifies. In this case, the intensities of the MQ coherences can be calculated explicitly for arbitrary coupling constants $b_{ij}$ [21]. The model accurately describes [22,23] the evolution of the first few experimental MQ intensities in a cubic lattice [24] and pseudo-1D spin chain [25]. In experiments, yy-zz average Hamiltonian ($a = 0$, $b = 1$, and $c = -1$) has been used, and $c = 2^{1/2}$ has been used in the zz-model to match the strength of the local fields. At long times, the model predicts [26] for concentrated spin systems

$$m_2(t) = t^2 M_2 , \qquad (28)$$

where $M_2 = \sum b_j^2/4$ is the conventional second moment of the absorption line. For dilute spin systems, $m_2(t) = t / T_2$, where $T_2$ is the decay time of the free induction signal. The reason why the growth of $m_2(t)$ in eq.(28) is quadratic in time and not linear, as it would be expected for the weak irreversibility, is that the zz Hamiltonian preserves individual z-components of spins. Therefore, the correlation functions (25) do not decay to zero, and the correlation times (26) do not exist.

*b) 1D spin chain with nearest-neighbors interactions.*

For a 1D spin chain with yy-zz Hamiltonian and equal nearest-neighbors interactions only, the MQ intensities oscillate between 0Q and 2Q [27]. No higher-order MQ coherences are generated. Therefore, $m_2(t)$ is limited: $m_2(t) \leq 4$ and the system has ideal reversibility, when the observable is $S_x$ and the perturbation is a uniform x-rotation of all spins. Inclusion of long-range interactions beyond the nearest neighbors spoils this ideal reversibility. The model is an interesting example demonstrating that an infinite growth of correlations between spins does not lead to irreversibility unless these correlations have needed transformation properties with respect to the perturbation Hamiltonian.

## VIII. FINITE CLUSTERS

For a cluster of $N$ spins ½, there are $2^N$ integrals of motion, which are the diagonal elements of the density matrix in a frame where the Hamiltonian is diagonal. Each pair of degenerate levels adds one more integral of motion. The projection of the density matrix on the subspace of the integrals is conserved and, therefore, the evolution is non-ergodic. The example of such non-ergodic behavior has been studied by simulating the process of spin diffusion in spin chains [28,29]. It has been found that the polarization of the initially polarized spin remains higher than the equilibrium value $1/N$ at all times and that spin diffusion fails to bring the system to equilibrium. In finite clusters $m_2(t)$ is limited by $m_2(t) \leq N^2$, so the dynamics is also non-mixing. Non-ergodic evolution makes the time-averaged values of the observables to be different from the equilibrium values, while the absence of mixing creates irregular oscillations around the average values. Such behavior has been observed experimentally for spin diffusion in a ring of six dipolar-coupled nuclear spins [30]. The experimental dynamics was in close agreement with the calculation performed for this system by J. S. Waugh.

One would expect that with increasing size of the spin cluster the effect on the dynamics of the exact integrals will decrease, because the number of the elements of the density matrix $2^{2N}$ grows much faster than the number of integrals $2^N$. Such decrease of the role of exact integrals at increasing cluster size has been demonstrated in simulations [31]. In the thermodynamic limit $N \to \infty$ the existence of the "microscopic" integrals of motion becomes unimportant. The only integrals of motion which impose explicit limitations on the dynamics are the additive integrals associated with global symmetries. As an example the Hamiltonian (1) with $a = b$ is invariant under z-rotations. As a result, in addition to energy, $S_z$ is conserved. An assumption that, within these limitations, the system is fully thermalized leads to the two-temperature thermodynamic theory [32,33] which has been very successful in describing various phenomena in solid-state NMR.

Dynamics of small spin clusters can be handled by numerical simulations. A direct diagonalization of the Hamiltonian can be applied to $N \sim 15$ spins ½ [34]. In the simulation [9] for $N = 25$ spins ½, an estimation of traces [35] has been used. Even though the number of quantum energy levels is huge in such clusters, the systems are still too small to adequately reproduce irreversibility of either the Loschmidt or the partial echo. As it follows from our discussion, one can only hope to get the initial part of the evolution with $m_2(t) < N$. As an example, 25-spin cluster will adequately represent a larger system at initial times, when only the nQ coherences with n = 0, 2, and 4 are present. This is not enough to predict the asymptotic long-time behavior of $m_2(t)$ in macroscopic systems.

## IX. PARTIAL ECHO

The Hamiltonian (1) is invariant under a uniform π-rotation of all spins. This global discrete symmetry makes each of the energy levels doubly degenerate. If ψ is an eigenstate of the Hamiltonian $H_{dz}$ and $S_z$, $exp(iS_x)ψ$ is also an eigenstate with the same energy. One can chose the eigenstates to be even and odd superpositions $ψ_{g,u} = ψ \pm exp(iS_x)ψ$. $S_x$ has no matrix elements between the states of different parity, and its evolution happens independently in the subspaces of even and odd states. In large spin systems, this has no observable consequences unless the symmetry is broken. If Δ is a small perturbation, linear in spins and $[H,Δ] \neq 0$, it creates a slow evolution between the subspaces. Δ can be a distribution of resonance frequencies (chemical shifts in NMR). We can note that the symmetry breaking has been also used in liquid-state NMR to access the long-lived singlet state of a spin pair [36] or a three-spin state [37] which are less sensitive to spin-lattice relaxation. The partial echo of small amplitude [38,39] is created by the Hahn's spin-echo pulse sequence $(π/2)_y – τ – π_x$ [40]. The first pulse creates the initial state, then, a free evolution follows, and subsequent π-pulse changes the sign of Δ without affecting the main Hamiltonian $H$. The total Hamiltonian changes from $H + Δ$ to $H – Δ$, and after another period of evolution τ, the echo is formed. Compared to the Loschmidt echo, the partial echo does not require the change of sign of the entire Hamiltonian and the need to use an approximate average Hamiltonian to accomplish it. The amplitude of the partial echo can be estimated [38,39] as $A_e \approx |\Delta|/|H|$, and its decay time $T_e \approx |\Delta|^{-1}$. In NMR experiments, the echo decay time $T_e$ can be 3-4 orders of magnitude longer than the characteristic time of spin dynamics $T_2$.

The Loschmidt echo physically reconstructs the initial state of the system, but it can be also viewed as the method of recovering the *information* about the initial state after a period of free evolution. The partial echo does not return the system to its initial state. In fact, the density matrix at the center of the echo is very different from the initial density matrix (the echo shape is different from the initial free induction signal). However, the partial echo performs the same task of reconstructing the information about the initial state. From a practical point of view, the partial echo greatly expands the time frame, at which such information recovery can be accomplished.

The time scale can be expanded even further in the suspended echo experiment [41] which uses the Hahn's stimulated echo pulse sequence [40]. The information about the initial state is stored during a long suspension time, and then retrieved by application of a single "reading" pulse. The information storage time is practically limited only by the spin-lattice relaxation time $T_1$, which can be extremely long for nuclear spins ½ in crystals, especially at low temperatures of the lattice [42,43]. As it has been stated in [42], $T_1$ can be "astronomic". In a room-temperature experiment [41] the echo in naphthalene has been recovered after a suspension time seven orders of magnitude longer than the characteristic time of spin dynamics $T_2$.

The existence of the partial echo allows excitation of sharp NMR signals in systems with dipolar-broadened spectra. Such signals can be used in MRI [44,45] and diffusion measurements [46].

## X. DISCUSSION

The purpose of science is predicting future. For a dynamical system of interest, the extent to which the past can be reconstructed, or the future predicted, depends on how deterministic is its evolution. For large classical systems, the determinism is limited to predicting the values of the additive integrals of motion and their densities. As an example, a trajectory of a macroscopic object can be predicted only because the total momentum of its atoms is conserved. In a similar way, the densities of additive integrals are the slow deterministic variables in hydrodynamics. Determinism and reversibility are the two sides of the same question if we formulate it in the following way. Suppose that at the moment $t = 0$ the value of the observable of interest (expectation value in quantum mechanics) is $M(0)$. Can we reconstruct an information about this value from the measurements done at the later moment $t = \tau$, after a period of free evolution? The Loschmidt echo experiment (its time-reversed part) can be viewed as a measuring procedure which performs this task. The rate, at which the echo is spoiled, or the growth rate of $\delta M(\tau) = M(0) - M(2\tau)$ can be used to quantify the irreversibility. In classical systems, $\delta M(\tau)$ grows exponentially for the majority of initial states (more accurate statements, including the possibilities of long-lived correlations can be found in the book [47]). Quantum systems have better reversibility. In some cases, one can even expect weak irreversibility with $\delta M(\tau) \propto \tau$.

There is another important difference between quantum and classical systems. While mixing and irreversibility in a classical system are mostly dictated by its Hamiltonian, in quantum systems all three factors are important: 1) the main Hamiltonian which governs the dynamics, 2) the observable of interest, and 3) the Hamiltonian of perturbation. As we have seen, the use of the transformation properties of the density matrix and the main Hamiltonian, with respect to the perturbation Hamiltonian, allows calculating the decay of the Loschmidt echo and formulating the criteria of irreversibility.

From a practical perspective, the beauty of the "magic echo" [1] is that it turned a purely theoretical concept of the Loschmidt echo into the experimental reality and demonstrated a possibility of recovering the information about the observable, which is not an integral of motion. Despite good reversibility of spin dynamics, the time scale of the "magic echo" is short. The reason is that there are no known ways of accurately reversing the sign of the Hamiltonian. The Hamiltonian with reversed sign is created as the lowest-order average Hamiltonian, while in multi-spin systems, the contribution to the dynamics of the higher-order terms of the average Hamiltonian is not small [48].

The discovery of the partial echo [38,39] offered a practical alternative to the Loschmidt echo. Instead of (inaccurate) reversal of the sign of the entire main Hamiltonian, one can precisely reverse the sign of a small perturbation with simpler structure. The result can be a small-amplitude partial echo with extremely long lifetime. In terms of recovering the information about an initial state, the partial echo does the same job as the Loschmidt echo. The information about the initial state can be stored for even longer times in the "suspended echo" experiment [41] and retrieved by an application of a single "reading" pulse.

The fundamental importance of the problem of reversibility of quantum dynamics is contrasted by a slow progress in this field. The reasons are abundant. The arsenal of theoretical methods to analyze correlated long-time dynamics of many-body quantum systems is limited. Exact solutions are helpful, but there are only few, and they cover very special cases. Direct numerical simulations can only handle the clusters which are too small to reproduce long-time behavior of macroscopic systems.

With a limited input from theory and simulations, the role of experimental studies increases. Nuclear spins ½ in crystals are probably the best experimental objects. Spin Hamiltonians in this case are known with very high accuracy, and spin degrees of freedom can be extremely well isolated from the lattice. Systems of coupled spins consist of the simplest quantum objects, and the question is whether spin dynamics is sufficiently rich and representative to allow generalization of the results. We think it is. As one of examples of bridging nuclear spin dynamics to the dynamics of other systems we can mention the behavior at high spin polarizations. In this case, spin dynamics is the dynamics of a low-density Bose-gas of magnons [49,50]. The behavior of a spin system is similar to that of, say, cold gases and includes the phenomena like Bose-condensation [50,51]. Theoretical analysis of spin dynamics in this case is simpler because only a subset of quantum states is involved. Unfortunately, the experiments at low spin temperatures [52] are challenging, and it is not an active area of research at present.

Development of NMR instrumentation has been driven by important applications in chemistry and biology. As a result of this development, modern NMR spectrometers are advanced tools capable of revealing very fine details of spin dynamics. Solid-state NMR may continue to add empirical pieces to our still fragmentary knowledge of the fundamentals of many-body quantum dynamics.

## ACKNOWLEDGEMENT

This paper is written in memory of John S. Waugh and his contribution to the field of spin dynamics. A stimulating discussion with James Yesinowski is greatly appreciated.